\documentclass[fleqn,usenatbib]{mnras}

\usepackage{newtxtext,newtxmath}
\usepackage{color}

\usepackage[T1]{fontenc}
\usepackage{ae,aecompl}

\usepackage{graphicx}	% Including figure files
\usepackage{amsmath}	% Advanced maths commands
\title[Black Hole Spin in the AGN channel]{Constraining the LIGO/Virgo AGN channel with black hole spins}

\author[B.McKernan \& K.E.S. Ford]{B. McKernan$^{1,2,3,4}$\thanks{E-mail:bmckernan at amnh.org (BMcK)} \& K.E.S. Ford$^{1,2,3,4}$
\\
$^{1}$Department of Astrophysics, American Museum of Natural History, New York, NY 10024, USA\\
$^{2}$Center for Computational Astrophysics, Flatiron Institute, New York, NY 10010, USA\\
$^{3}$Graduate Center, City University of New York, 365 5th Avenue, New York, NY 10016, USA\\
$^{4}$Department of Science, BMCC, City University of New York, New York, NY 10007, USA\\
}

\date{Accepted XXX. Received YYY; in original form ZZZ}

\pubyear{2023}

\begin{document}
\label{firstpage}
\pagerange{\pageref{firstpage}--\pageref{lastpage}}
\maketitle

\begin{abstract}
Merging black holes (BH) are expected to produce remnants with large dimensionless spin parameters ($a_{\rm spin} \sim 0.7$). However, gravitational wave (GW) observations with LIGO/Virgo suggest that merging BH are consistent with modestly positive but not high spin ($a_{\rm spin} \sim 0.2$), causing tension with models suggesting that high mass mergers are produced by hierarchical merger channels. Some BH also show evidence for strong in-plane spin components. Here we point out that 
\emph{spin down} of BH due to eccentric prograde post-merger orbits within the gas of an active galactic nucleus (AGN) disk can yield BH with masses in the upper mass gap, but only modestly positive $a_{\rm spin}$, and thus observations of BH with low spin \emph{do not} rule out hierarchical models. We also point out that the fraction of BBH mergers with significant in-plane spin components is a strong test of interactions between disk binary black holes (BBH) and nuclear spheroid orbiters. Spin magnitude and spin tilt constraints  from LIGO/Virgo observations of BBH are an excellent test of dynamics of black holes in AGN disks, disk properties and the nuclear clusters interacting with AGN.

\end{abstract}


%%%%%%%%%%%%%%%%%%%%%%%%%%%%%%%
\section{Introduction}
Binary black hole (BBH) mergers observed in gravitational waves (GW) with LIGO-Virgo can originate from multiple channels, including from the death of massive binary stars \citep[e.g.][]{Belczynski10,deMink16}, or from BH that pair-up dynamically after formation \citep[e.g.][]{Antonini14,Rodriquez16a,Fragione19}. A promising dynamics channel is BBH mergers in AGN disks \citep[e.g.][see also \citet{Arca23} for a recent review]{McK14,Bartos17,Stone17}. Broad expectations for this channel include: efficient IMBH ($>100M_{\odot}$) formation \citep[e.g.][]{McK12,Bellovary16,Yang19,Secunda19,Tagawa20}, occasional asymmetric mass mergers \citep{McK20b,Tagawa20} and possibly residual orbital eccentricity in the LIGO band \citep{Samsing22}. 

Black hole (BH) masses can be used to discriminate  between merger channels. For example, since massive stars are not believed to directly produce BH $\sim 50-120 M_{\odot}$, GW detections of progenitor BH in this 'upper mass' gap are suggestive of BH that are themselves merger products \citep[e.g.][]{Gerosa19,Hiromichimassgap21,Ford22,Gayathri23}. Likewise an observed pile-up of BH at $\sim 40M_{\odot}$ \citep{O3aspins} might be associated with the lower end of the expected upper mass gap. The fact that the global peak of the mass distribution for BH involved in BBH mergers is at low mass (around $10M_{\odot}$) is another fascinating clue. Since neutron stars ($\sim 1.4M_{\odot}$) natal kicks are observed $\leq 800{\rm km/s}$, relatively low mass BH (e.g. $\sim 10M_{\odot}$) are likely formed with modest non-zero kicks $\leq 110{\rm km/s}(M_{\rm NS}/1.4M_{\odot})(M_{\rm BH}/10M_{\odot})^{-1}$ \citep{Coleman22}. A deep potential well could help retain lighter kicked BH in order to promote their subsequent merger, driving up the merger rate at generally low BH mass.%Random supernova kicks towards a binary partner might harden a fraction of binaries, but equally a (moderately) deep potential well could help retain an overdensity of such BH in order to promote their subsequent merger as observed. (!!!Help me out here with this logic?!!!)

Black hole spins can also provide important clues to BBH channel origins \citep[e.g.][]{Zevin20,Chase20, Galaudage21} as well as specifically testing AGN channel models \citep{Avi22}. The effective spin ($\chi_{\rm eff}$) distribution observed by LIGO/Virgo is biased to positive values \citep{O3aspins}, which is not expected for spherically symmetric dynamical models (i.e. mergers in clusters). Observed spins are also not uniformly aligned with each other and the BBH orbital angular momentum, which naively might be expected in some stellar origin scenarios. Rather, the spin distribution observed has some dynamical characteristics, including some weight at $\chi_{\rm eff}<0$ and some events with evidence for strong in-plane spin components, but with an overall symmetry-breaking (positive) bias. Some BBH have also been observed to possess strong in-plane spin components \citep{Vijay22}. Also observed is a fascinating anti-correlation between $\chi_{\rm eff}$ and mass ratio ($q$) in BBH mergers \citep{Callister21}. Such a anti-correlation is hard to generate in both field and dynamical channels. However, AGN can yield such an effect if more massive BH spend longer (and spin-up) in AGN disks and if there is a bias against retrograde BBH \citep[e.g.][]{McK22, Wang21}. \citet{Santini23} also find such an effect emerges straightforwardly  assuming prograde BBH mergers happen at an AGN disk migration trap.

Here we focus on BH spin as a test of models of the AGN channel. In particular, we discuss why hierarchical mergers in AGN can (eventually) result in BH with low spin and why observations of BH with low spin \emph{do not} rule out a hierarchical origin. We also briefly discuss the implications of  strong in-plane spin components for the AGN channel. Finally, we discuss the implications of the observed spin distributions for models of AGN disks and the dynamics and accretion history of the embedded populations within them.

\section{Black hole spin}
It is still unclear what spins BH are born with. Observations of BH within our own Galaxy indicate moderate to high ($a \sim 0.3 - 0.9$) BH spins \citep{Chris19}, but these BH accrete from X-ray binary companions, and so do not provide an unbiased diagnostic of BH spin at birth \citep[e.g.][]{MayaVicky22}. It has been proposed that BH are born with very low spin $a \sim 0.01$ \citep{Fuller19}. However BH in LIGO observations have BH spin magnitudes typically an order of magnitude larger than this \citep{O3aspins}.

Depending on the initial BH spin, accretion onto the BH after birth can alter the spin magnitude and/or torque the BH spin alignment. Figure~\ref{fig:acc_mode} illustrates the effect of accretion direction on spin magnitude (length of spin vector) and orientation (direction of spin vector), assuming locally disk-like accretion geometry. Prograde accretion (top panel) increases an initially positive spin magnitude ($a_{\rm spin}$) and torques spin orientation towards the orbital angular momentum of the accretion flow. Retrograde accretion (bottom panel) decreases $a_{\rm spin} \rightarrow 0$ and then $a_{\rm spin}<0$ and drives orientation towards anti-alignment. The timescale of this process depends on the rate of accretion. \citet{Bogdanovic07} find that BH can be torqued into alignment with a large-scale gas flow once $\sim 1-10\%$ of the BH mass has been accreted. Since the Eddington mass doubling rate is $\sim 40{\rm Myr}$, a period of $\sim [1,{\rm few}]$Myr of accretion at the Eddington rate should be sufficient to torque BH spins into alignment with a massive accretion flow. Thus, a population of embedded objects in AGN should be biased towards positive spin alignments, depending on how long the AGN disk persists. 

There is much more certainty about the spins of merging BH. Numerical relativity results show that much of the BBH orbital angular momentum at merger goes into the spin of the resulting merged BH. Thus, the merged spin ($a_{\rm merged}$) can be written as \citep[e.g.][]{Tichy08}
\begin{eqnarray}
    a_{\rm merged} &\approx& 0.686(5.04\nu -4.16\nu^{2}) \\
    &+& 0.4\left(\frac{a_{1}}{(0.632 +1/q_{\rm bin})^{2}}+\frac{a_{2}}{(0.632+q_{\rm bin})^{2}} \right)
\end{eqnarray}
where $q_{\rm bin}=M_{2}/M_{1}$ is the binary mass ratio, $\nu=q_{\rm bin}/((1+q_{\rm bin})^{2})$ is the symmetric mass ratio and $a_{1},a_{2}$ are the binary component spin parameters. For most cases, with modest spins, moderate mass ratios ($q_{\rm bin} \sim 1$), $a_{\rm merged} \sim 0.7$.

A population of merging BH that include the products of prior mergers would therefore be expected to have spins $a \sim 0.7$. However, as we shall point out below, the role of gas accretion and torquing changes this basic conclusion, at least in the AGN channel.

\section{Damping timescales}
\label{sec:damp}
Gas in AGN disks acts to damp prograde orbital eccentricity \citep[e.g.][]{McK12} but pumps orbital eccentricities of retrograde orbits \citep{Secunda21}.
For prograde orbits in protoplanetary disks with small orbital eccentricity ($e<2h$), $e$ decays exponentially over time $\tau_{e} \approx h^{2}\tau_{\rm mig}$, where $\tau_{\rm mig}$ is the migration timescale,  $h=H/r$ is the disk aspect ratio with $H$ the disk scale height and $r$ the radius of the orbiter in the disk \citep{Papaloizou00}. At larger eccentricities, decay goes as $\dot{e} \propto e^{-2}$ \citep{Bitsch10}. 

Rescaling the disk damping timescale $t_{\rm damp}$ \citep{Tanaka04} to an  embedded BH in an AGN disk we find
\begin{equation}
    t_{\rm damp}=\frac{M_{\rm SMBH}^{2}h^{4}}{m_{\rm BH}\Sigma a^{2} \Omega}
\end{equation}
where $M_{\rm SMBH}$ is the supermassive black hole (SMBH) mass, $m_{\rm BH}$ is the embedded black hole mass, $\Sigma$ is the disk surface density, $a$ is now the orbital semi-major axis and $\Omega$ the Keplerian orbital frequency. The strong dependence on the aspect ratio of the disk ($h^{4}$) in $t_{\rm damp}$ implies we should expect prohibitively long orbital damping times either in the outer cooler disk or in a puffed up, hot inner disk. Since damped circularized orbits will preferentially form BBH in dynamical encounters \citep{Secunda21,Rowan22,Jairu23,Stan23}, this suggests that BBH formation is more likely during encounters in the thinnest, densest regions of AGN disks.

We can usefully parameterize $t_{\rm damp}$ as
\begin{equation}
    t_{\rm damp} \sim 0.1{\rm Myr} \left(\frac{q}{10^{-7}}\right)^{-1}\left(\frac{h}{0.03}\right)^{4}\left(\frac{\Sigma}{10^{5}{\rm kg m^{-2}}}\right)^{-1}\left(\frac{a}{10^{4}r_{g}}\right)^{-1/2}
\label{eq:damp}
\end{equation}
where $q=m_{\rm BH}/M_{\rm SMBH}$ is the mass ratio of the embedded BH to the SMBH, $\Sigma \sim 10^{5}{\rm kg m^{-2}}$ is a surface density consistent with  moderately dense outer regions ($a \sim 10^{4}r_{g}$) of a \citet{Sirko03} model AGN disk, where $r_{g}=GM_{\rm SMBH}/c^{2}$ is the gravitational radius. Note that more massive BH (larger $q$) have orbits damped faster. This is important, since it implies that more massive BH in the AGN channel should on average spend more time spinning up and torquing into alignment with the disk gas than less massive BH. Such an effect could also help explain the bias towards positive $\chi_{\rm eff}$ observed in the more massive component of BBH \citep{Callister21}. 

From eqn.~(\ref{eq:damp}), in the cool outer part of this disk model ($>3 \times 10^{4}r_{g}$), $t_{\rm damp}$ is long $\geq 1{\rm Myr}(q/10^{-7})^{-1}$. However, at the thinnest part of this disk model $t_{\rm damp}$ is very short
\begin{equation}
    t_{\rm damp} \sim {\rm kyr}\left(\frac{q}{10^{-7}}\right)^{-1}\left(\frac{h}{0.01}\right)^{4}\left(\frac{\Sigma}{10^{7}{\rm kg m^{-2}}}\right)^{-1}\left(\frac{a}{10^{3}r_{g}}\right)^{-1/2}.
\end{equation}
But in the radiation-pressure dominated innermost disk, the disk puffs up again and 
\begin{equation}
t_{\rm damp} \sim 0.6{\rm Myr}\left(\frac{q}{10^{-7}}\right)^{-1}\left(\frac{h}{0.05}\right)^{4}\left(\frac{\Sigma}{10^{6}{\rm kg m^{-2}}}\right)^{-1}\left(\frac{a}{10^{2}r_{g}}\right)^{-1/2}.
\end{equation}
Thus, orbital eccentricity damping is most efficient in the thinnest parts of the  \citet{Sirko03} model between $\sim [10^{2},10^{4}]r_{g}$. In the \citet{TQM05} disk model, we also find damping timescales are either very short $t_{\rm damp} \sim {\rm kyr}$ in the very thin mid-disk region ($h/R \sim 10^{-3}$) and prohbitively long $t_{\rm damp}>{\rm Myr}$ otherwise (since $h/R \sim 0.05$ in both the inner and outer disk regions).

With $t_{\rm damp}$, the scaling timescale for eccentricity damping, we can now estimate how long orbital damping can take in these disk models. At small initial orbital eccentricity ($e_{0}<2h$), assuming exponential decay \citep{Papaloizou00} 
\begin{equation}
    e(t)=e_{0}\rm{exp}(-t/t_{\rm damp})
\end{equation}
and so within $\sim 2-3 t_{\rm damp}$, $e_{0}$ is damped to approximately circular ($e <0.01$) \footnote{assuming there are no additional orbital perturbations from dynamical encounters.}. From eqn.~\ref{eq:damp} if $e_{0} \sim 0.06$ at $a \sim 10^{4}r_{g}$ in a \citet{Sirko03} disk, the eccentricity is damped by gas to $e<0.01$ within $\sim 0.5$Myr.

At large eccentricities ($e>2h$), and assuming orbital inclination is negligible, we can use the approximation of \citep{Horn12} 
\begin{equation}
    t_{e} \sim \frac{t_{\rm damp}}{0.78} \left[ 1 -0.14 (e/h)^{2} + 0.06 (e/h)^{3} \right]
\label{eq:te}
\end{equation}
to find the timescale on which large orbital eccentricity becomes damped. The last term in eqn.~{\ref{eq:te}} dominates if $e/h \gg 7/3$, i.e. across most of the disk models above for $e>0.1$. 

For a thermal distribution of initital eccentricities, as the result of an equipartition of energy in a relaxed (either wholly or in part) nuclear star cluster, we expect an orbital distribution function 
\begin{equation}
    f(e)de = 2e de
\end{equation}
such that the median eccentricity is $e \sim 1/\sqrt{2} \sim 0.7$ and there is a uniform probability distribution of $e^{2}$. Thus, at modest disk thickness $h \sim 0.05$, $t_{e} \sim 175 t_{\rm damp}(e_{0}/0.7)^{3}$ or $t_{e} \sim 0.1{\rm Myr} (q/10^{-7})^{-1}(e_{0}/0.7)^{3}$ at the thinnest part of the \citet{Sirko03} disk
 model around $a \sim 10^{3}r_{g}$.

The regions of AGN disks where orbits are most rapidly circularized (i.e. regions with small values of $h$) will be where the relative energy of BH encounters is small enough that binary formation is efficient \citep[e.g.][]{Yaping22,Rowan22,Jairu23,Stan23}. Thus, we expect new BBH to predominantly form in AGN disks in the thin mid-disk region. The subsequent migration of such BBH either inwards or outwards, away from this disk region, will make dynamical encounters with eccentric orbiters more likely. Such encounters will be capable of either hardening, softening/ionizing the BBH, depending on the details of the encounter \citep[e.g.][]{Leigh18,Wang21,Adam22}. Note also $a \sim 10^{3}r_{g}$ is a plausible location for a migration trap \citep{Bellovary16} in such a disk, although \citet{Grishin23} suggest that such traps occur at radii $\times 3-5$ further out in the disk.

\section{Retrograde accretion: spin-down of eccentric orbiters}
The direction of gas flow onto objects embedded in disks is a function of orbital eccentricity \citep[e.g.][]{Bailey21,Yaping22}.  Physically, embedded orbiters on nearly circular orbits experience inflow into their Hill sphere from co-orbital gas leading to prograde accretion via mini-disks. If the embedded orbiters have  eccentric orbits, Keplerian shear leads to retrograde inflow overcoming the prograde circum-single disk. Figure~\ref{fig:circ_ecc} shows a cartoon sketch of these two modes. 

\begin{figure}
\begin{center}
\includegraphics[width=0.85\linewidth,angle=0]{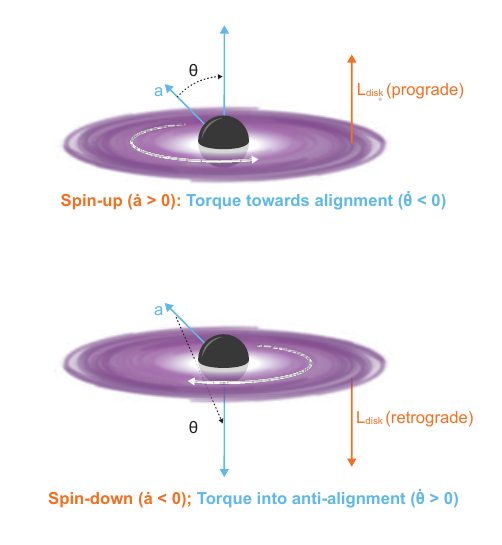}
\end{center}
\caption[BrokenSym]{Cartoon illustrating prograde and retrograde accretion onto BH embedded in an AGN. Top panel shows a gas minidisk (with prograde orbital angular momentum $L_{\rm disk}$) accreting onto a BH. Blue vector labelled $a$ corresponds to an initial BH spin vector mis-aligned with the accretion flow. Dashed line shows the direction of torque of the BH spin over time, through decreasing angle $\theta$ towards alignment with mini-disk, but also increasing spin magnitude (longer final blue vector parallel to $L_{\rm disk}$). Bottom panel is similar except the accretion minidisk has retrograde orbital angular momentum. BH spin at first \emph{decreases} in magnitude towards $a=0$ (vanishing vector) and then grows increasingly negative ($a<0$) over time approaching full anti-alignment with the greater AGN disk (unlabelled downward pointing final spin vector).
\label{fig:acc_mode}}
\end{figure}

\begin{figure*}
\begin{center}
\includegraphics[width=0.85\linewidth,angle=0]{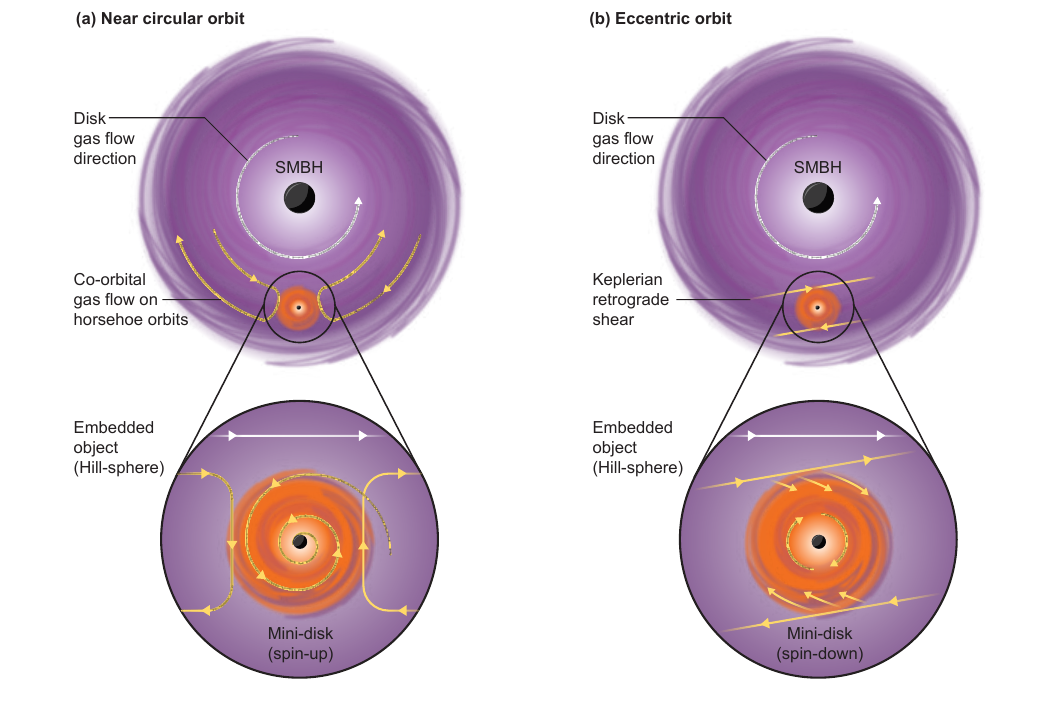}
\end{center}
\caption[BrokenSym]{(a) Top panel: Cartoon of accretion onto a BH embedded in an AGN disk on an approximately circular orbit. White arrow indicates AGN gas flow direction and direction of orbit of the embedded BH. Yellow arrows indicate the relative flow of gas in the frame of the embedded BH. (a) Bottom panel: Zoom in to Hill sphere of embedded BH shows prograde nature of accretion onto the BH and therefore spin-up and torquing into eventual alignment with the AGN disk. (b) Top panel: As in (a) but embedded BH is on an eccentric orbit and Keplerian shear dominates over co-orbital gas flow. (b) Bottom panel: Keplerian shear in the frame of the embedded BH, leads to retrograde accretion and therefore spin-down and torquing into eventual anti-alignment with the AGN disk.
\label{fig:circ_ecc}}
\end{figure*}

Fig.~\ref{fig:circ_ecc} (a) shows a top-down view of a near circular embedded BH orbiter. Arrows indicate the flow of gas in the disk (white) and in the frame of the orbiter (yellow). Gas flow on horseshoe orbits relative to the embedded orbiter is apparent, leading to (bottom left, a zoom-in of the orbiters' Hill sphere) a prograde flow of gas onto the embedded object, leading to spin-up (increasing the magnitude of $a$) and eventual torquing of spin into alignment with the AGN disk. Fig.~\ref{fig:circ_ecc}(b) shows a top-down view of an eccentric orbiter. Horseshoe orbits of gas are no longer apparent and the background gas flow exhibits a Kepelerian retrograde shear. The bottom panel of Fig.~\ref{fig:acc_mode} shows a zoom-in on the orbiters' Hill sphere indicating retrograde accretion, leading to spin-down (decreasing the magnitude of $a$) and eventual torquing of spin into anti-alignment with the AGN disk. 

Recently \citet{YiXian22} have demonstrated  a bifurcation in accretion direction on orbiters embedded in gas disks. Orbiters on nearly circular orbits accrete from prograde mini-disks within their Hill sphere, whereas embedded orbiters on eccentric orbits above a transition value ($e_{t}$) accrete from \emph{retrograde} mini-disks, where $e_{t}$ is \citep{YiXian22}
\begin{equation}
    e_{t} > h \sqrt{(1+\lambda^{2})\rm{max}[1,3^{1/3}(q^{1/3}/h)^{2}] -1}
\end{equation}
 with $\lambda \sim 1.3$ a numerical constant, $q_{t}=q/h^{3}$ is the thermal  mass ratio and $h$ is the disk scale-height. Effectively the bifurcation corresponds to $e_{t} \geq \lambda h$ for $q_{t} \leq 1$ (sub-thermal orbits) and $e_{t} \geq (\lambda h) q_{t}^{1/3}$ for $q_{t}>1$. In a \citet{Sirko03} model disk, all orbits are sub-thermal ($q_{t}<1$) for masses $<10^{2}M_{\odot}$ and super-thermal only in the thinnest regions of the disk for IMBH ($>10^{2}M_{\odot}$). In a \citet{TQM05} model disk, all orbits are super-thermal at the thinnest part of the disk, and sub-thermal elswehere. As a rule of thumb therefore, if BH orbital eccentricity is roughly $>\times 1.3$ the disk scale height of a \citet{Sirko03} disk and most scale heights of a \citet{TQM05} disk, for sub-IMBH masses, it will accrete retrograde. Thus, BH orbital eccentricities $e>0.08$ on average ($\overline{h} \sim 0.05$) in a \citet{Sirko03} disk should drive retrograde accretion.

\begin{figure}
\begin{center}
\includegraphics[width=0.85\linewidth,angle=0]{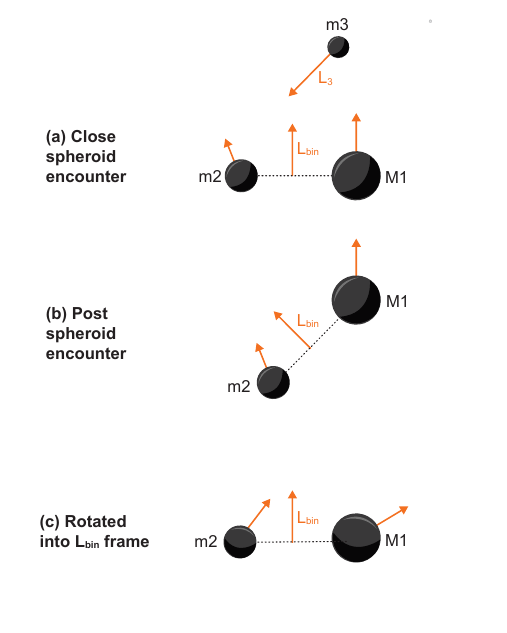}
\end{center}
\caption[BrokenSym]{Cartoon illustrating how a large $\chi_{\rm p}$ component can arise in the AGN channel. See also \citep{Samsing22} for a comparable  illustration. ($M_{1},m_{2}$) is a binary embedded in the AGN disk, with orbital angular momentum ($L_{\rm bin}$) aligned with that of the AGN disk ($L_{\rm disk}$). The spin of the heavier primary ($M_{1}$) is aligned with the disk, and that of the secondary ($m_{2}$) is mis-aligned. Such a binary arrangement is consistent with the ($q,\chi_{\rm eff}$) correlation in \citet{Callister21}. Also depicted in (a) is a near-miss encounter by tertiary ($m_{3}$) from the nuclear spheroid population, with mis-aligned orbital angular momentum ($L_{3}$). Conservation of orbital angular momentum, leads to (b), where the BBH has been ejected from the disk with new (resultant) orbital angular momentum. Note however that the spins of the individual BH have not been torqued and remain oriented as in (a). (c) shows (b) but rotated so $L_{\rm bin}$ is now vertical, showing clearly a strong spin component in-plane. 
\label{fig:chip}}
\end{figure}

\section{Post-merger kicks}
The anisotropic emission of GW from a merging BBH means that the merged BH will recoil with a kick from the merger site \citep[e.g.][and references therein]{Centrella10}. Maximum merger kick velocity is $v_{\rm kick}\sim 5000{\rm km/s}$ \citep{Campanelli07,Gonzalez07} for approximately equal mass merging BH, with maximal and anti-aligned spins. Kick velocity drops considerably as the mass ratio of the merging BBH ($q_{BBH}=M_{2}/M_{1}$) decreases since $v_{\rm kick} \propto q^{2}_{BBH}$ \citep{Centrella10}.

Keplerian orbital velocities at the thinnest regions of the \citet{Sirko03} disk model ($a \sim 10^{3-4}r_{g}$) span $\sim 3-10 \times 10^{3}{\rm km/s}$. For a BBH merger in this region of the disk, kicks $v_{\rm kick}\geq 30-100{\rm km/s}$ will generate orbital eccentricities $e>0.01$. Thus, it is straightforward to generate eccentric orbits of merged BBH in AGN disks. Spin mis-alignments can drive larger kicks ($\sim 10^{3}{\rm km/s}$), as inferred from some LIGO BBH observations \citep{Vijaykick}. Since we generally expect spin mis-alignment from random sortings in the AGN channel, the likelihood of large $v_{\rm kick}$ in this channel is quite high. As a result, a sizeable fraction of merged BBH in the AGN channel should have modest $e \sim [0.02,0.5]$ post-merger. 

From \S\ref{sec:damp} above, in a \citet{Sirko03} disk model, we expect the gas disk will dampen orbital eccentricity to near circular in $\leq 0.5$Myr in the thinnest regions of this disk. If $1-10\%$ mass accretion is sufficient to torque 
 a BH into alignment with the disk \citep{Bogdanovic07}, then assuming Eddington-limited accretion, in a few Myr, a BH could be torqued from fully aligned ($a_{\rm merged} \sim 0.7$) into anti-alignment, via spin-down and driving spins to negative magnitudes ($a_{\rm merged} \sim -0.9$). In $\leq 0.5$Myr therefore, significant spin-down from $a_{\rm merged} \sim 0.7$ can occur. Differences in average spin magnitudes ($a_{\rm spin}$) for BH with masses in the upper mass gap $\geq 50M_{\odot}$ (presumed to result from hierarchical mergers) and low mass BH (say $\sim 10M_{\odot}$) can help constrain this phenomenon in AGN disks.

The fact that negative spins are not preferred among LIGO BBH observations, suggests for the AGN channel both that: (i)  retrograde BBH are disfavored (avoiding the formation of a negative spin BH) \citep{Wang21,McK22,Santini23} \footnote{Retrograde BBH experience eccentricity pumping and torquing towards disk alignment (LANL group; private communication, see also \citet{Lubow15}).} and (ii) spin-down does not progress so far that by the time BBH form, most BH are not  
negative spin. This implies damping in the regions BBH formation and merger occurs must be very efficient and therefore since $t_{\rm damp} \propto h^{4}$, these regions of the AGN disks must be geometrically thin. 

\section{$\chi_{\rm p}$ in the AGN channel}
$\chi_{\rm eff}$ is the projection of the mass-weighted spins of a BBH onto the binary orbital angular momentum ($L_{\rm bin}$) around its center of mass. $\chi_{\rm p}$ is the effective precession spin given by 
\begin{equation}
\chi_{\rm p} = {\rm max} \left[a_{1,\perp},\frac{q(4q+3)}{4+3q}a_{2,\perp} \right]
\end{equation}
where $a_{i,\perp}$ is the component of spin perpendicular to the direction of $L_{\rm bin}$ \citep{o3b}. 
Several mergers observed by LIGO may have significantly non-zero $\chi_{\rm p}$ \citep{Vijay22}, which appears to suggest a dynamical origin. Significant in-plane spin was first highlighted for the AGN channel by \citet{Tagawa20} and \citet{Samsing22} indicating what could be a 'smoking gun' for an origin from the AGN channel.

In the AGN channel, the most straightforward way of generating a strong $\chi_{\rm p}$ component for BBH is to start with a primary BH spin that is relatively well aligned with the AGN disk and a secondary with random (modestly positive) BH spin. Such a configuration is appropriate for generating a ($q,\chi_{\rm eff}$) anti-correlation as discussed in \citet{Callister21, McK22}. Figure~\ref{fig:chip}(a) depicts this 
 binary set-up. Now introduce a close pass by a nuclear cluster object on a disk-crossing orbit (object $m_{3}$, with inclined orbital angular momentum ($L_{3}$) also depicted in Fig.~\ref{fig:chip}(a)). Such an interaction conserves orbital angular momentum so the BBH orbital angular momentum ($L_{\rm bin}$) tilts to a new (resultant) BBH plane orientation and the BBH is kicked out of the disk. Fig.~\ref{fig:chip}(b) shows the newly inclined BBH and Fig.~\ref{fig:chip}(c) shows panel (b) rotated into the frame of $L_{\rm bin}$ (so that $L_{\rm bin}$ is now vertical) to illustrate the strong spin components in the BBH plane.

Several conditions apply to this AGN channel scenario. First, there must be a close interaction with a spheroid orbiter. The cross-section for encounters with a spheroid orbiter of radius $R_{\ast}$ and mass $M_{\ast}$ is \citet{Leigh18} \begin{equation}
    \Gamma_{\rm NSC} \approx \sigma \rho  \left(\frac{R_{\ast}^{2}}{M_{\ast}}\right)\left[1+\left(\frac{v_{\rm esc}}{\sigma}\right)^{2}\right]
\end{equation}
where $\rho \sim M_{\rm NSC}/R_{\rm NSC}^{3}$ is the spheroidal volume density of NSC objects, $\sigma \sim \sigma_{0} +(GM_{\rm SMBH}/r)^{1/2}$ is the velocity dispersion of the NSC, and $v_{\rm esc}$ is the escape velocity from the BBH. For encounters at moderate disk radius, Keplerian dispersion dominates and this implies that an interaction such as in Fig.~\ref{fig:chip}(a) most likely occurs in the innermost regions of the AGN disk, since the cross-section for interaction is highest there (e.g. Fig.~1 of \citet{Leigh18}). 
Note that over a long AGN lifetime much of the spheroid component (particularly in inner regions, where disk-crossing is frequent) can be captured by the disk \citep{Fabj20,MacLeodLin20,Syeda23}. Thus, such an interaction must occur early on in an AGN lifetime ($<$Myr).

Second, this encounter must harden the binary since the BBH must persist to merger. The binary separation at the moment of encounter must therefore be \citep{Leigh18}
\begin{equation}
    a_{\rm bin}<12^{1/3}R_{H} \left(\frac{\mu_{\rm bin}}{m_{3}} \right)^{1/3}
\end{equation}
where $R_{H}=a(q/3)^{1/3}$ is the BBH Hill radius, with $a$ the BBH semi-major axis and $q=M_{\rm bin}/M_{\rm SMBH}$ the BBH mass ratio, and $\mu_{\rm bin}=M_{1}M_{2}/M_{\rm bin}$ is the binary reduced mass. Third, the BBH must also merge \emph{before} it is recaptured by the disk \citep[e.g.][emphasize this point]{Tagawa20}, because the orbital angular momentum of the binary will be rapidly realigned with the angular momentum of the disk, once the binary is recaptured, returning us to the configuration of Fig.~\ref{fig:chip}(a). A merger outside the disk could have a prompt EM counterpart, rather than a delayed EM counterpart as envisaged in \citep{Graham20a}, if the BBH has dragged gas with it as it passes through the AGN disk on each orbit (or it could lack any EM counterpart for lack of surrounding matter).

\section{Discussion}
\label{sec:discussion}
The AGN channel is fundamentally a dynamical channel with broken spherical symmetry, which yields hierarchical mass mergers more frequently (via retention of kicked BBH merger remnants) than any other LIGO channel. A BBH merger should naturally yield a high spin remnant, leading to the general expectation of seeing highly spinning, high mass BH progenitors for any hierarchical merger channel. However, in the AGN channel, we show here that the BH merger product is unlikely to retain that high spin for long. In particular, we point out that as long as the kick at merger yields a modest orbital eccentricity, BH produced in dynamical mergers in AGN disks must spin down (via retrograde accretion driven by Keplerian shear) while the BH orbital eccentricity is damped over time by disk gas. Going forward, the density of BH in AGN and the time between encounters (and therefore typical migration torques and timescales), as well as orbital damping timescales (a function of AGN gas density and scale height) should be constrained using Monte Carlo studies of this effect. 

An additional measurable parameter, $\chi_{\rm p}$, has an expected distribution for standard dynamical channels, but AGN again distort this expectation due to separate populations and the unique disk symmetry where many BBH may easily form. A large $\chi_{\rm p}$ component in a BBH merger can occur if the primary BH initially has spin strongly aligned with the AGN disk, but pre-merger the BBH experiences a close encounter with a spheroid orbiter (substantially inclined with respect to the orbit of the BBH center of mass of around the SMBH), likely in the inner disk. The result is a BBH kicked out of the AGN disk (see Fig.~\ref{fig:chip}) with new orbital angular momentum direction, but un-torqued spin components, leading to significant in-plane spin. Such mergers are an excellent test of the rate of encounters between the spheroid and disk components of nuclear star clusters interacting with AGN. Again, Monte Carlo studies of this encounter type can strongly constrain the disk and spheroid population in  unresolved nuclear star clusters hosting AGN. 

Negative $\chi_{\rm eff}$ mergers are disfavoured in the AGN channel for several reasons: 1) BBH with $L_{\rm bin}<0$ (retrograde orbit around their center of mass) are preferentially ionized or softened by moderately close tertiary encounters compared to prograde BBH \citep{Wang21}. 2) Retrograde BBH are also likely to have their eccentricity pumped by gas and therefore spend more time on average at wider separations \citep{Dong23}, making them more likely to be softened or ionized in tertiary encounters. 3) Retrograde BBH with $L_{\rm bin}$ not identically anti-aligned with $L_{\rm disk}$ will experience an accretion torque towards alignment with disk gas over time, flipping $L_{\rm bin}$ positive \citep{Lubow15}. 

Nevertheless, some $\chi_{\rm eff}<0$ mergers should happen occasionally in AGN disks \citep{McK22}. In order to preserve a negative $\chi_{\rm eff}$ binary to merger against dynamical prograde encounters, such a BBH is more likely to be massive, with $q \sim 1$. Negative spin in a BBH in the AGN channel can occur if there is a capture, or exchange on close pass, between a BH on a long-lived eccentric orbit and a BH on a nearly circular orbit, and the semimajor axis at binary formation is small enough that the GW merger timescale is shorter than the timescale to torque $L_{\rm bin}$ into alignment with the disk. The long-lived eccentric orbits required to form such BBH could persist in either the colder outer disk, or the puffed-up hot inner disk. In the colder outer disk, the rate of tertiary (potentially ionizing) encounters is significantly lower, so $\chi_{\rm eff}<0$ BBH are more likely to surive to merger in that region.

\section{Conclusions}
Spin information from LIGO BBH merger observations provides important clues to the underlying population and merger channel details. In particular, for the AGN channel, spin constraints (both $\chi_{\rm eff}$ and $\chi_{\rm p}$ distributions and correlations) are strongly constraining on the details of how gas and dynamics drive mergers of BH embedded in AGN gas disks. 

Here we point out that the AGN channel can produce hierarchical merger products that should \emph{spin-down} over time. Average BBH spin parameter estimates from LIGO-Virgo will help constrain the gas damping timescale in AGN and therefore $\rho_{\rm disk}, h$ as well as typical merger locations. We also point out that BBH mergers with large $\chi_{\rm p}$ are straightforward to produce in the AGN channel due to interactions between a BBH and a non-disk component tertiary (a point also made by \citet{Tagawa20} and \citet{Samsing22}). 

{\section{Acknowledgements.}} BM \& KESF are supported by NSF AST-2206096 and NSF AST-1831415 and Simons Foundation Grant 533845 as well as Simons Foundation sabbatical support. The Flatiron Institute is supported by the Simons Foundation. Thanks to Lucy Reading-Ikkanda for her excellent illustrations.

\section*{Data Availability}
Any data used in this analysis are available on reasonable request from the first author (BM).

\bibliographystyle{mnras}
\bibliography{refs} % if your bibtex file is called example.bib

\end{document}